\documentclass{blois}

\input{journalabbvr.texinc}

\def\be{\begin{equation}}
\def\ee{\end{equation}}
\def\bea{\begin{eqnarray}}
\def\eea{\end{eqnarray}}

\newcommand{\Photo}{}

\begin{document}
\vspace*{4cm}
\title{Probing solutions to the $S_8$ tension with galaxy clustering}

\author{Pedro Carrilho$^a$, Chiara Moretti$^{b,c,a}$, Maria Tsedrik$^a$}

\address{$^a$ Institute for Astronomy, The University of Edinburgh, Royal Observatory, Edinburgh EH9 3HJ, UK}

\address{$^b$ SISSA - International School for Advanced Studies, Via
  Bonomea 265, 34136 Trieste, Italy}

\address{$^c$ Centro Nazionale ``High Performance Computer, Big Data and Quantum Computing''}

\maketitle

\abstracts{
The current discrepancy between the CMB and weak lensing measurements of the amplitude of matter fluctuations, the so-called $S_8$ tension, has attracted a great deal of recent attention, as it may show a crack in the $\Lambda$CDM model of cosmology. We review the evidence for this tension and describe potential solutions, focusing on extensions of the standard cosmological model, including interacting dark energy and modified gravity. We present a likelihood analysis of the BOSS DR12 data, probing these alternative models as well as $\Lambda$CDM. From this analysis, we show hints of non-standard cosmology compatible with those seen in weak lensing observations, demonstrating that interacting dark energy or modified gravity can explain them successfully. We then discuss the robustness of these results to analysis choices, as well as future paths to confirm them with additional data and further distinguish between models.}

\section{Introduction}

The $\Lambda$CDM model of cosmology has enjoyed incredible success in describing the Universe, fitting the increasingly precise data from the cosmic microwave background (CMB)~\cite{planck2018cosmo} and from surveys of the large-scale structure (LSS). However, this model is unsatisfactory from the theoretical point of view, given that the nature of the dark sector is not known at a fundamental level, with Cold Dark Matter (CDM) being know only to be a non-relativistic species without non-gravitational interactions and dark energy being simply parameterised via the cosmological constant, $\Lambda$. Additionally, possible cracks in this model could be appearing, as recent precise measurements of the late-time Universe are finding results that are in disagreement with those of the early Universe~\cite{verde2019}, which could represent new windows into the physics of the dark sector. 

One of these so-called cosmic tensions relates to the amplitude of density fluctuations at late time. Direct measurements from weak lensing surveys~\cite{DES:2021bvc,DES:2021vln,KiDS:2020suj,Li:2023tui,Dalal:2023olq,DES_KiDS} find consistently lower values of the amplitude parameter $S_8\equiv \sigma_8\sqrt{\Omega_m/0.3}$ than that obtained by the Planck analysis of CMB data under the assumption of the $\Lambda$CDM model~\cite{planck2018cosmo}. While some aspects of the nonlinear modelling still require some clarification,~\cite{Arico:2023ocu,DES_KiDS} this tension has generated substantial interest in alternatives to $\Lambda$CDM that can lower the amplitude, ranging from models of the dark sector and its interactions,\cite{Simpson:2010vh,Pourtsidou:2013nha,Skordis:2015yra,Lesgourgues:2015wza} to extensions of General Relativity.\cite{Clifton:2011jh,Nguyen:2023fip,Moretti:2023drg}

Galaxy clustering is a powerful tool for investigating alternatives to $\Lambda$CDM that can explain the $S_8$ tension, particularly since measurements of the growth rate, $f$, give an alternative probe of their effects. This is exactly what we do here, showing our analyses of the BOSS data in Section~\ref{sec:Results} for the two extensions to $\Lambda$CDM described in Section~\ref{sec:Theory} as well as for the standard model. We then show forecasts for stage-IV spectroscopic surveys such as ESA's Euclid satellite mission~\cite{laureijs2011}, and the Dark Energy Spectroscopic Instrument (DESI)~\cite{desi2016} in Section~\ref{sec:Forecasts}.

\section{Theoretical solutions for the $S_8$ tension}
\label{sec:Theory}

A solution to the $S_8$ tension requires a novel effect that suppresses the growth of structure at late times. One broad class of models that do this adds interactions to dark matter, typically with dark energy. This interaction is selected to slow the growth of dark matter fluctuations, thus resolving the tension. A promising model that does just that is the \emph{Dark Scattering} model,\cite{Simpson:2010vh,Baldi:2014ica,Baldi:2016zom} in which dark matter and dark energy interact, exchanging momentum via elastic scattering, but not transferring energy. In this model, which we label $wA$CDM, the only equation that is modified is the Euler equation for dark matter, given by
\begin{equation}
    \theta_{\rm DM}'+(\mathcal{H}+A a \rho_{\rm DE})\theta_{\rm DM}+\nabla^2\phi=0,
\end{equation}
where $\theta$ is the velocity divergence, $\phi$ is the gravitational potential, $\rho_{\rm DE}$ is the dark energy density, assumed to be a perfect fluid with equation of state $w$, $\mathcal{H}=a'/a$ is the conformal Hubble rate, with $a$ the scale factor and a prime denoting conformal time derivatives. The interaction strength $A$ is defined as $A\equiv(1+w)\sigma_D/m_{\rm DM}$, and depends on the interaction cross section $\sigma_D$ and the dark matter mass $m_{\rm DM}$. Additionally, its dependence on $w$ selects its sign to be equal to that of $1+w$. It is clear that if $A$ is positive, the interaction acts as an additional friction force on the dark matter fluid, reducing growth at late times when dark energy becomes relevant.

Alternatively, lowering the strength of gravity at late times also reduces the growth of structure compared to $\Lambda$CDM and can in principle resolve the $S_8$ tension.
We investigate here a particular functional form for the growth rate $f\equiv d\log \delta/d\log a$, given by the well-known growth index $\gamma$, which can be used to parameterise various modified gravity models~\cite{linder2007}
\begin{equation}
    f(a)=\Omega_m(a)^{\gamma}\,,
\end{equation}
with $\gamma$ assumed to be constant and reducing to $\Lambda$CDM for the value $\gamma\approx0.545$, while larger $\gamma$ corresponds to slower growth at late times. We also add the sum of neutrino masses, $M_\nu$, as a free parameter to this model. This improves its realism, and allows us to investigate how the constraints change, since increasing the mass of neutrinos also has the effect of lowering the amplitude of perturbations, albeit in a scale-dependent way.

Both of these specific alternatives can explain the observed low value of $S_8$, by an appropriate choice of their parameters, $A$ for Dark Scattering and $\gamma$ for modified gravity. To distinguish between them one can take advantage of their different growth history, as opposed to only the current value of the amplitude of fluctuations in the form of $S_8$. For this reason, spectroscopic galaxy clustering is the ideal way to probe these models, as it directly measures the rate of change of $\sigma_8$, via its measurement of $f\sigma_8\approx d\sigma_8/d\log a$. In the following section we show our data analyses of the BOSS DR12 dataset for these two models.

\begin{figure}[h]
\begin{minipage}{0.495\linewidth}
\centerline{\includegraphics[width=1\linewidth]{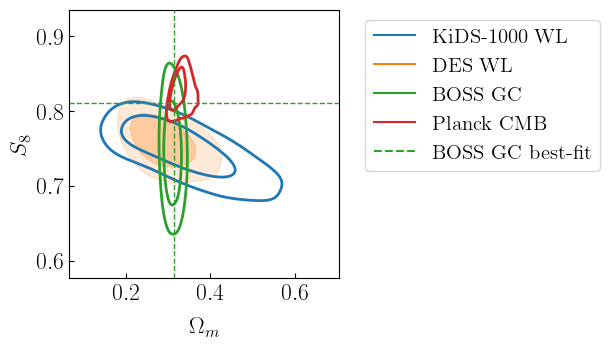}}
\end{minipage}
\begin{minipage}{0.495\linewidth}
\centerline{\includegraphics[width=1\linewidth]{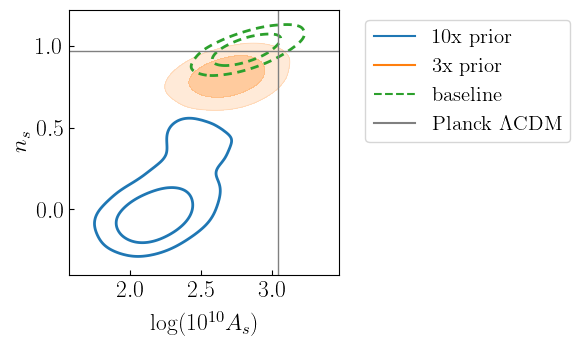}}
\end{minipage}
\caption[]{Left: Comparison of the constraints on $S_8$ and $\Omega_m$ from the BOSS analysis in $\Lambda$CDM to the weak lensing and Planck constraints, showing also the best-fit point for the BOSS analysis. Right: Variation of the constraints on $A_s$ and $n_s$ when nuisance parameter priors are broadened by 3 and 10 times relative to the baseline case, demonstrating the informative nature of these priors and the importance of prior volume effects.}
\label{fig:LCDM}
\end{figure}

\section{BOSS analyses}
\label{sec:Results}

\begin{figure}
\begin{minipage}{0.49\linewidth}
\centerline{\includegraphics[width=0.92\linewidth]{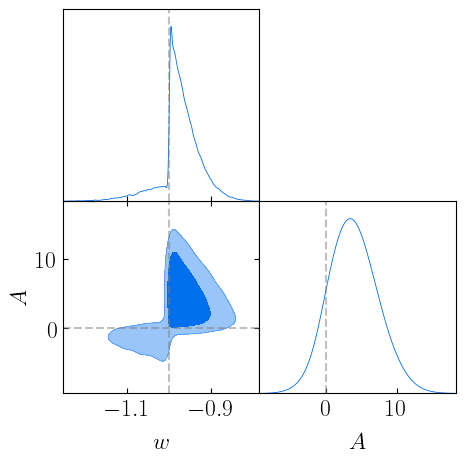}}
\end{minipage}
\hfill
\begin{minipage}{0.49\linewidth}
\centerline{\includegraphics[width=0.92\linewidth]{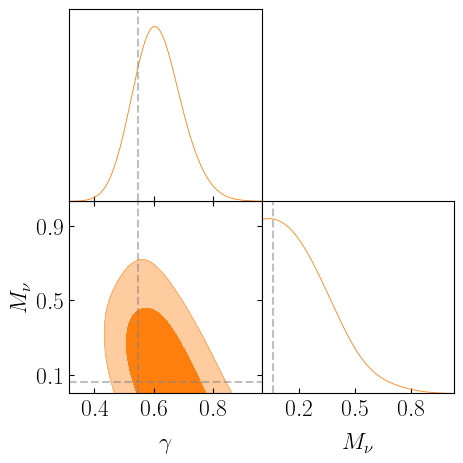}}
\end{minipage}
\caption[]{BOSS constraints on beyond-$\Lambda$CDM models, using a CMB prior on the primordial amplitude parameters $A_s$ and $n_s$. Left: $wA$CDM dark energy parameters. Right: $\gamma\nu$CDM parameters.}
\label{fig:wA_gamma_mnu}
\end{figure}

We now show our analyses of BOSS DR12 with the two models described above. This data is comprised of power spectra at redshifts $z=0.38$ and $z=0.61$ in two sky patches, which we analyse up to $k_{\rm max}=0.2~h/{\rm Mpc}$. We also add measurements of the BAO scale at redshifts $z=0.106,~0.15, ~0.38,~ 0.61,~2.334$, and use a Big Bang Nucleosynthesis prior on the baryon density of $100\omega_b=2.268\pm0.038$. To allow for constraining the parameters of the extensions, we also use CMB information in the form of a $3\sigma$ Planck prior on the parameters of the primordial power spectrum of $\log 10^{10} A_s=3.044\pm0.42$ and $n_s=0.9649\pm0.012$.

We model the galaxy power spectrum in redshift space using the effective field theory of large-scale structure (EFTofLSS)~\cite{Baumann:2010tm,Carrasco:2012cv}. We use a particular model that has been previously applied to BOSS,\cite{ivanov2020b,chudaykin2020,philcox2022} but use the new code PBJ~\cite{moretti2023}. 
This model includes 11 nuisance parameters per redshift and sky, including 4 bias parameters, 3 shot noise parameters and 4 counter-term parameters.
More details on the definitions and the full structure of the model can be found in Carrilho et al.\cite{carrilho2021,carrilho2023} and Moretti et al.\cite{Moretti:2023drg} Priors for nuisance parameters are set according to the so-called ``east coast" prescription~\cite{philcox2022} in the baseline analyses but are also varied to test how they influence the result. This model is employed in a fast MCMC analysis pipeline, using analytical marginalisation over 8 of the 11 nuisance parameters and an emulator for the linear matter power spectrum~\cite{arico2021}. We show results from this pipeline for 3 different models: $\Lambda$CDM, Dark Scattering, labeled $wA$CDM, and for the growth index parameterisation, labeled $\gamma\nu$CDM, which also includes massive neutrinos.

We begin with our results for $\Lambda$CDM, obtained in Carrilho et al.~\cite{carrilho2023} In the left panel of Fig.~\ref{fig:LCDM} we show our constraints on the $S_8$-$\Omega_m$ plane, comparing with weak lensing and CMB measurements. We see that the BOSS contour agrees very well with weak lensing measurements, with its mean value being $S_8=0.746$, in slight tension with Planck and in broad agreement with EFT-based BOSS analyses.\cite{damico2020,ivanov2020b,philcox2022} However, the best-fit value of $S_8=0.810$ is in far greater agreement with Planck, already indicating that the posterior is highly non-Gaussian. This is further demonstrated in the right panel of Fig.~\ref{fig:LCDM}, where we see the dependence of the results on the size of the priors of the nuisance parameters. This indicates the existence of a prior volume effect,\cite{Hadzhiyska:2023wae} which lowers the mean value of $A_s$/$\sigma_8$. One must thus be careful when drawing conclusions on the $S_8$ tension from these analyses. This effect has also been seen in other works~\cite{simon2022} and some solutions have been recently proposed using Jeffreys priors~\cite{Donald-McCann:2023kpx} or profile likelihoods.~\cite{Moretti:2023drg,Holm:2023laa}

We now move to our results for $wA$CDM and $\gamma\nu$CDM. When using only galaxy clustering data, there is a strong degeneracy between the primordial amplitude $A_s$ and the extension parameters $A$ or $\gamma$. We do not show those results here, but the interested reader can find them in Carrilho et al.\cite{carrilho2023} and Moretti et al.\cite{Moretti:2023drg} We show instead only the results using additional CMB information in Fig~\ref{fig:wA_gamma_mnu}. We can see that for both models there is a small preference of $\sim1\sigma$ for the extensions. For $wA$CDM we find $w=-0.972^{+0.036}_{-0.029}$ and $A=3.9^{+3.2}_{-3.7}~b/{\rm GeV}$, while for $\gamma\nu$CDM we get $\gamma=0.612^{+0.075}_{-0.090}$ and $M_\nu<0.3~{\rm eV}$ at 68 \% CL. Both cases show good agreement with weak lensing results, having $S_8\approx0.78$, as well as with the Planck constraints on the other cosmological parameters and therefore both are able to resolve the $S_8$ tension, without there being a preference for either in the BOSS data. In the right panel of Fig.~\ref{fig:wA_gamma_mnu} a slight degeneracy can be seen between $\gamma$ and $M_\nu$, and our analysis with fixed $M_\nu$ gives instead $\gamma=0.647\pm0.085$.
 
\section{Forecasts for stage-IV surveys}
\label{sec:Forecasts}

Constraints improve further by adding more observables or novel datasets. Here we show that explicitly in both cases, generating forecasts for $wA$CDM including the bispectrum for a Euclid-like spectroscopic survey; and for $\gamma\nu$CDM in a DESI-like configuration. 

\begin{figure}
\begin{minipage}{0.49\linewidth}
\centerline{\includegraphics[height=0.62\linewidth]{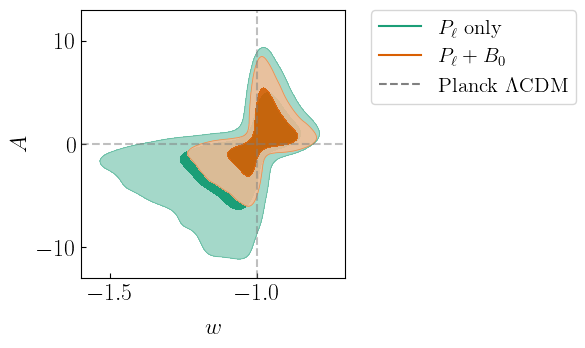}}
\end{minipage}
\hfill
\begin{minipage}{0.49\linewidth}
\centerline{\includegraphics[height=0.62\linewidth]{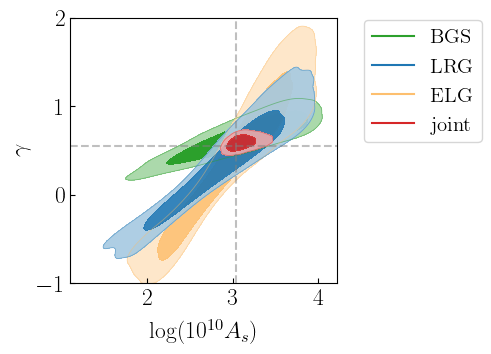}}
\end{minipage}
\caption[]{Results for the forecasts for stage-IV surveys. Left: Constraining power for $wA$CDM for a Euclid-like spectroscopic survey for power spectrum-only ($P_\ell$) analyses vs power spectrum + bispectrum ($P_\ell+B_0$) analyses. Right: Constraining power for $\gamma\nu$CDM for a DESI like survey, comparing the results for each of its three galaxy samples (BGS, ELG and LRG) with the joint analysis, showing explicitly the breaking of the $A_s-\gamma$ degeneracy.}
\label{fig:forecasts}
\end{figure}

In the left panel of Fig.~\ref{fig:forecasts}, we show the improvement of constraints that can be achieved when adding bispectrum data to the analysis.\cite{tsedrik2023} The bispectrum is the simplest source of non-Gaussian information and due to its distinct dependence on the bias parameters from the power spectrum, it can improve constraints on the $wA$CDM parameters by 30\% with respect to the power spectrum-only analysis.

In the right panel of Fig.~\ref{fig:forecasts}, we show that combining the different DESI samples at multiple redshifts can improve constraints on the $\gamma$ parameter, even without adding CMB information, resulting in an uncertainty $\sigma_\gamma=0.058$, 30\% better than the BOSS result. With Planck information the improvement is instead of 45\%. We can therefore conclude that the higher precision of DESI and its broad range of samples will greatly improve constraints on alternatives to $\Lambda$CDM.

\section{Conclusions}
Multiple models can rectify the $S_8$ tension, and it is important to use galaxy clustering to distinguish them. Our analyses of the BOSS data show that concordance can be re-established within both $wA$CDM or $\gamma\nu$CDM, and we place constraints on both. We find that neither is currently preferred. Additionally, we see that priors are informative and change results so care must be taken with their interpretation.
Our forecasts for stage-IV spectroscopic surveys show substantial improvements from the addition of the bispectrum and from multi-$z$ analyses, demonstrating great promise in the future for measuring extensions to $\Lambda$CDM.

\section*{Acknowledgments}

We thank Alkistis Pourtsidou for her guidance during the development of these projects. PC's research is supported by grant RF/ERE/221061. CM's work is supported by the Fondazione ICSC, Spoke 3 Astrophysics and Cosmos Observations, National Recovery and Resilience Plan (Piano Nazionale di Ripresa e Resilienza, PNRR) Project ID CN\_00000013 ``Italian Research Center on High-Performance Computing, Big Data and Quantum Computing'' funded by MUR Missione 4 Componente 2 Investimento 1.4: Potenziamento strutture di ricerca e creazione di ``campioni nazionali di R\&S (M4C2-19)" - Next Generation EU (NGEU).  MT's research is supported by a doctoral studentship in the School of Physics and Astronomy, University of Edinburgh.

\bibliography{biblio.bib}

\end{document}